\documentclass[12pt]{article}
\usepackage{amsmath}
\usepackage{graphicx}
\usepackage{natbib}
\usepackage{url}
\usepackage{blkarray}

\usepackage[utf8]{inputenc}
\usepackage{amsthm}
\usepackage{natbib}
\usepackage[depth=2]{bookmark}
\usepackage[ruled,vlined]{algorithm2e}
\usepackage{setspace}
\usepackage{amsmath}
\usepackage[english]{babel}
\usepackage{caption}
\usepackage{enumerate}
\usepackage{enumitem}
\usepackage{comment}
\usepackage{bm}
\usepackage{cleveref}
\usepackage{tikz}
\usepackage{dsfont}
\usepackage{amsfonts}
\usepackage{amssymb}
\usepackage{accents}
\usepackage{multirow}
\usepackage{rotating}
\usepackage{pdflscape}
\usepackage{hyperref}
\usepackage{lmodern}
\usepackage{siunitx}
\usepackage{booktabs}
\usepackage{etoolbox}
\usepackage{caption}
\usepackage[T1]{fontenc}

\usepackage{mathtools}

\DeclarePairedDelimiter\floor{\lfloor}{\rfloor}

\newtheorem{thm}{Theorem}
\newtheorem{cor}{Corollary}

\newtheorem{prop}{Proposition}

\newtheorem{assumption}{Assumption}

\theoremstyle{definition}
\newtheorem{definition}{Definition}

\newtheorem*{thm*}{Theorem}
\newtheorem*{cor*}{Corollary}
\newtheorem*{ex*}{Example}
\newtheorem*{prop*}{Proposition}
\newtheorem*{lemma*}{Lemma}
\newtheorem*{assumption*}{Assumption}

\theoremstyle{definition}
\newtheorem*{definition*}{Definition}
\newtheorem*{remark*}{Remark}

\newcommand{\mt}{\top}

\newcommand{\norm}[2]{\left\lVert#1\right\rVert_{#2}}

\newcommand{\ind}[1]{\mathds{1}\left(#1\right)}
\newcommand{\bzero}{{\mathbf{0}}}		

\newcommand{\pdset}[1]{\cS^{#1 \times #1}_{+}}

\newcommand{\bhat}[1]{\widehat{#1}}
\newlength{\dhatheight}

\newcommand{\pare}[3]{\left#1 #2\right#3}

\newcommand{\bv}{\bm{v}}

\newcommand{\Sbb}{\mathbf{S}}

\newcommand{\Ar}{\mathrm{A}}
\newcommand{\Br}{\mathrm{B}}

\newcommand{\Er}{\mathrm{E}}
\newcommand{\Fr}{\mathrm{F}}

\newcommand{\Mr}{\mathrm{M}}

\newcommand{\Prr}{\mathrm{P}}

\newcommand{\Rr}{\mathrm{R}}
\newcommand{\Srr}{\mathrm{S}}

\newcommand{\Ur}{\mathrm{U}}

\newcommand{\bX}{\bm{X}}

\newcommand{\bmu}{\bm{\mu}}

\newcommand{\bSigma}{\bm{\Sigma}}

\newcommand{\bXi}{\bm{\Xi}}

\newcommand{\bbR}{\mathbb{R}}

\newcommand{\bbN}{\mathbb{N}}

\newcommand{\bbI}{\mathbb{I}}

\newcommand{\cE}{\mathcal{E}}

\newcommand{\cG}{\mathcal{G}}
\newcommand{\cH}{\mathcal{H}}
\newcommand{\cI}{\mathcal{I}}

\newcommand{\cP}{\mathcal{P}}

\newcommand{\cS}{\mathcal{S}}
\newcommand{\cT}{\mathcal{T}}

\newcommand{\brank}{\textbf{rank}}

\newcommand{\bdiag}{\textbf{diag}}	
\newcommand{\bspan}{\textbf{span}}	

\newcommand{\STEP}{\mathrm{2-step}}

\newcommand{\TPR}{\mathrm{TPR}}
\newcommand{\FPR}{\mathrm{FPR}}

\newcommand\indep{\protect\mathpalette{\protect\independenT}{\perp}}
\def\independenT#1#2{\mathrel{\rlap{$#1#2$}\mkern2mu{#1#2}}}	

\newcommand{\Unif}{\textbf{Uniform}}

\newcommand{\vertiii}[1]{{\left\vert\kern-0.25ex\left\vert\kern-0.25ex\left\vert #1 
		\right\vert\kern-0.25ex\right\vert\kern-0.25ex\right\vert}}

\setcitestyle{numbers,square}
\begin{document}

\def\spacingset#1{\renewcommand{\baselinestretch}
{#1}\small\normalsize} \spacingset{1}

{
	\title{\bf Hub Detection in Gaussian Graphical Models}
	\author{
		José Á. Sánchez Gómez \and  Weibin Mo \and  Junlong Zhao \and Yufeng Liu\thanks{ José Á. Sánchez Gómez is Assistant Professor, Department of Statistics, University of California Riverside, Riverside, CA 92521, USA. E-mail: \href{mailto:josesa@ucr.edu}{josesa@ucr.edu}. Weibin Mo is Assistant Professor, Mitchell E. Daniels, Jr. School of Business, Purdue University, West Lafayette, IN 47907, USA. Email: \href{mailto:harrymok@purdue.edu}{harrymok@purdue.edu}. Junlong Zhao is Professor, School of Statistics, Beijing Normal University, Beijing, China, 100875. E-mail: \href{mailto:zhaojunlong928@126.com}{zhaojunlong928@126.com}. Yufeng Liu is Professor, Department of Statistics and Operations Research, Department of Genetics, Department of Biostatistics, University of North Carolina at Chapel Hill, NC 27599, USA. E-mail: \href{mailto:yfliu@email.unc.edu}{yfliu@email.unc.edu}.\\
		Corresponding Authors: Yufeng Liu (\href{mailto:yfliu@email.unc.edu}{yfliu@email.unc.edu}), Junlong Zhao (\href{mailto:zhaojunlong928@126.com}{zhaojunlong928@126.com}).}
	}
	\date{}
	\maketitle
}

\bigskip
\begin{abstract}
	Graphical models are popular tools for exploring relationships among a set of variables. The Gaussian graphical model (GGM) is an important class of graphical models, where the conditional dependence among variables is represented by nodes and edges in a graph. In many real applications, we are interested in detecting hubs in graphical models, which refer to nodes with a significant higher degree of connectivity compared to non-hub nodes. A typical strategy for hub detection consists of estimating the graphical model, and then using the estimated graph to identify hubs. Despite its simplicity, the success of this strategy relies on the accuracy of the estimated graph. In this paper, we directly target on the estimation of hubs, without the need of estimating the graph. We establish a novel connection between the presence of hubs in a graphical model, and the spectral decomposition of the underlying covariance matrix. Based on this connection, we propose the method of {\it inverse principal components for hub detection} (IPC-HD). Both consistency and convergence rates are established for IPC-HD. Our simulation study demonstrates the superior performance and fast computation of the proposed method compared to existing methods in the literature in terms of hub detection. Our application to a prostate cancer gene expression dataset detects several hub genes with close connections to tumor development. 
\end{abstract}

\vfill

\newpage
\spacingset{1}
\setlength{\abovedisplayskip}{2pt}
\setlength{\belowdisplayskip}{2pt}

\section{Introduction}\label{section:introduction}

Research on graphical models has been prevalent in probability, statistics and machine learning  communities for many years \citep{lauritzen1996graphical,jordan1999introduction}. This line of research is based on the idea of representing relationships among variables through a network. For a graphical model, nodes correspond to variables in the model, and edges represent pairwise relationships between variables. In particular, for undirected graphical models, an edge between two nodes can indicate conditional dependence between the corresponding variables. For models of directed acyclic graphs, a directed edge may represent a causal relationship between the variables. A survey on graphical models, and their connections to statistics, probability and optimization can be found in \citet{wainwright2019high}.

Modeling the dependence of variables through a graphical model is very useful in practice. One example is the analysis of gene expression levels of cancer patients. Such datasets may consist of measurements of tens of thousands of gene segments for a set of patients. An important goal of modeling such data using graphical models is to estimate significant conditional dependencies among genes. Estimating these relationships can help understand how genes interact and function with each other. Examples of such analyses can be found in \citet{abegaz2013sparse,zhao2019cancer}.

Within the graphical model literature, the Gaussian Graphical Model (GGM) has received a lot of attention, partly due to its desirable statistical properties. It is known that the network structure of a GGM can be characterized by the sparsity pattern of the inverse covariance matrix, also known as the precision matrix \citep{wainwright2019high}. Due to the relationship between conditional independence and sparsity, a popular strategy is to estimate graphical models by solving a sparse penalized estimation problem. Examples include the neighborhood selection technique \citep{Meinshausen_2006}, the graphical LASSO (GLASSO) penalized likelihood approach \citep{yuan2007model}, and the restricted $\ell_1$-minimization method \citep{cai2011constrained}. 

In many applications, one is interested in detecting hubs in a graphical model, which correspond to groups of nodes with a higher degree of connectivity compared to other nodes. For example, in a gene expression dataset, a hub gene has a high degree of conditional dependence to other genes, so it may be pivotal at regulating specific genetic processes of interest \citep{BI201571}. In studies of political corruption, hubs may correspond to agents with a central role in corruption schemes \citep{10.1093/comnet/cny002}. In neurological applications, hubs can correspond to regions in the brain whose function is to interconnect multiple regions, and whose deterioration may be indicative of brain injury \citep{10.1093/brain/awv075}. In these applications, correct identification of hub variables is essential for understanding interactions among  variables of interest. 

Despite the importance of identifying hubs, only a few methods have been proposed in the literature for the estimation of hubs in GGMs. An early approach for hub estimation consists of thresholding the partial correlation matrix \citep{hero2011hub}. \citet{tan2014learning} proposed the Hub Graphical LASSO (HGL) which estimates hubs by decomposing the precision matrix into sparse and hub components. More recently, \citet{McGillivray_2020} proposed the hub-weighted GLASSO (HWGL) which fits the traditional GLASSO with adaptive weights that incorporate hub information. Note that both HGL and HWGL need to estimate the entire graphical model in order to identify its hubs. 

In this paper, we focus on the task of direct hub detection, without the need of estimating the graphical model. We first establish a notion of hubs in a precision matrix to represent the presence of a small set of variables with strong association to the remaining variables. Under mild conditions, we show that the existence of hub variables implies a low effective rank structure in the precision matrix. Furthermore, the entries of leading eigenvectors concentrate on the hub variables. As a consequence, we can identify hub variables from the top eigenvectors of the precision matrix. To our best knowledge, this is the first framework to connect the spectral decomposition of a precision matrix with the presence of hubs in a graphical model, although spectral methods have been extensively applied for a variety of contexts such as spiked covariance matrices \citep{baik2006eigenvalues,paul2007asymptotics}, principal component analysis \citep{hotelling1933analysis,chen2022distributed}, factor models \citep{fan2018robust}, and community detection in network data \citep{abbe2017community}. 

With this insight, we propose our {\it inverse principal components for hub detection method} (IPC-HD) for direct detection of hubs in GGMs. The proposed IPC-HD is shown to be effective without the need of the common sparsity assumption used in GGMs. By using tools of matrix perturbation theory and covariance concentration \citep{chen2020spectral, fan2018robust}, we provide performance guarantees for the IPC-HD when $n,p\to \infty$. We successfully show the consistency of our method under mild conditions of the sample size and effective dimension of the data. Our numerical study shows the superior performance and fast computation compared with several existing methods.

The rest of this paper is organized as follows. In Section \ref{section:framework}, we introduce the mathematical framework of GGMs, and establish the relationship between the presence of hubs in a precision matrix, and its spectral decomposition. We define our methodology, and give implementation details in Section \ref{section:methodology}. Theoretical properties of the IPC-HD are explored in Section \ref{section:theoretical}. Simulation studies are provided in Section \ref{section:simulations}. In Section \ref{section:real_data}, we consider an application of the IPC-HD to a prostate cancer gene-expression dataset. Some discussions of the findings and future directions can be found in Section \ref{section:discussion}.

\section{Gaussian Graphical Models and Hub Detection}\label{section:framework}

We establish some notational conventions. For two non-negative sequences $ \{a_{p}\}_{p=1}^{\infty} $ and $ \{ b_{p} \}_{p=1}^{\infty} $, we denote $ a_{p} = O(b_{p})$, or $ b_{p} = \Omega(a_{p}) $, if there exists $0 < C < +\infty$ such that $a_p \leq C\cdot b_p$ for all large $p$. Denote the asymptotic equivalence $a_{p}\propto b_{p}$ if $a_{p} = O(b_{p})$ and $b_{p} = O(a_{p})$. We denote $a_{p} = o(b_{p})$, or equivalently $b_p \gg a_p$, if ${a_{p}}/{b_{p}}\to 0$ as $p\to \infty$.
For $p\in\bbN$, we denote $\cP := \{1,2,\ldots, p\}$. Let $|A|$ denote the cardinality of a set $A$. Let $\pdset{p}$ be the set of $p\times p$ positive definite matrices. For $\Ar, \Br\in\pdset{p}$, we denote $\Ar \leq \Br$ if $\Br-\Ar\in\pdset{p}$, $\Ar_{\cdot k} = (A_{ik})_{i\in\cP}\in\bbR^{p}$ the $k$-th column vector of $\Ar$, and $\brank(\Ar)$ the rank of $\Ar$. We denote by $\|\cdot\|_2$ the $\ell_2$-norm on vectors, and by  $\vertiii{\cdot}_2$ the operator norm on matrices.

We consider the problem of learning the conditional dependence relationships among $p$ variables $\bX = \{X_k\}_{k\in \cP}$. We assume $\bX$ follows a multivariate normal distribution, \textit{i.e.}, $\bX\sim N_p(\bmu,\Sigma^{(p)})$, where $\bmu\in\bbR^p$, $\Sigma^{(p)} \in \pdset{p}$ is a positive definite \textit{covariance matrix}, and $\Theta^{(p)} = (\Sigma^{(p)})^{-1}$ is the corresponding \textit{precision matrix}. Let $  \cG := (\cP, \cE) $ be the network associated with $\bX$ such that $  (k, l) \notin \cE $ if and only if $ X_{k} \,\indep\, X_{l}\,|\,\bX_{\cP\backslash\{k, l\}} $ for all $ k, l \in \cP $. We refer to $\cG$ as the \textit{Gaussian graphical model for the distribution of $\bX$} \citep{wainwright2019high}. It is well known that, if $\bX$ is Gaussian, $ X_k \indep X_l \mid \bX_{\cP\setminus\{k,l\}} $ if and only if $ \Theta^{(p)}_{kl}=0$. From this, the graphical model $ \cG $ has the adjacency matrix $ \left[\ind{\Theta^{(p)}_{kl}\neq 0}\right]_{k, l \in \cP} $ \citep{wainwright2019high}.

\subsection{Motivation}\label{subsec:motivation}

\begin{figure}[h]
	\centering
	\includegraphics[width = 0.8\linewidth, height=4.7cm]{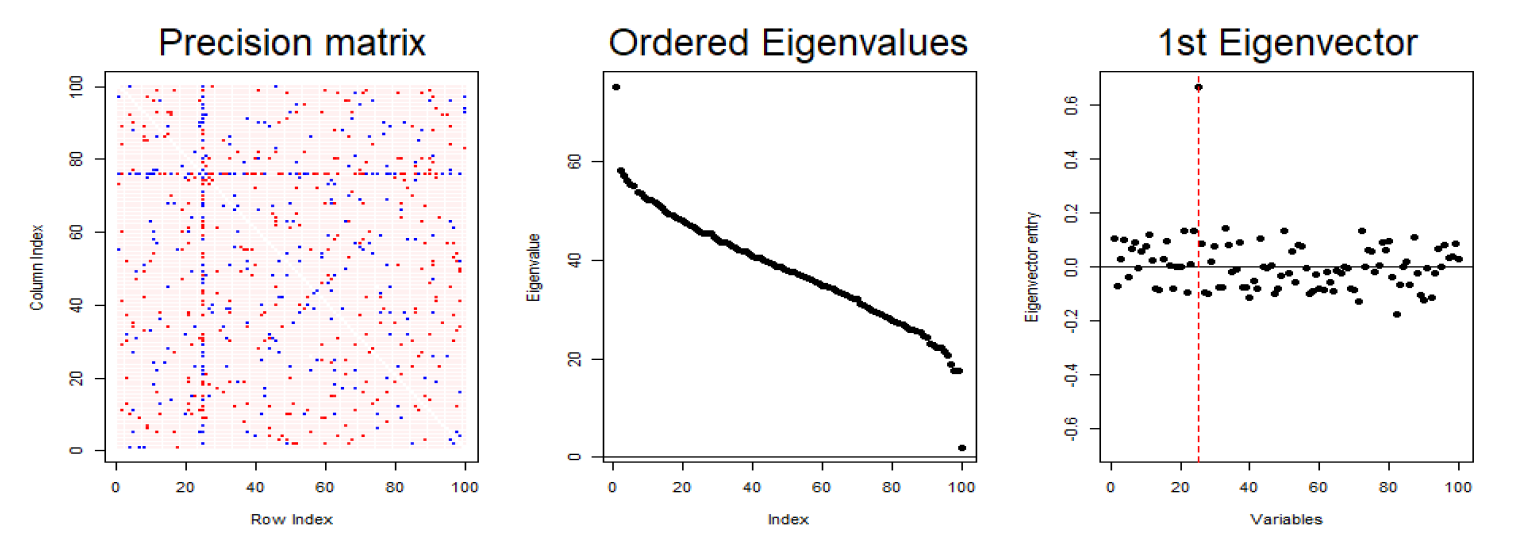}
	\caption{Plots for the illustrative example $\Theta^{(100)}_{1}$. Left panel: visualization of the precision matrix $\Theta^{(100)}_{1}$. The $25$-th variable is highly connected to the other variables. Center panel: plot of the ordered eigenvalues of $\Theta^{(100)}_{1}$. The first eigenvalue is separated from the rest. Right panel: coordinates of the first eigenvector of $\Theta^{(100)}_{1}$. The red line corresponds to the $ 25 $-th coordinate, which has a large magnitude compared to the rest.}
	\label{fig:quick_example}
\end{figure}

\begin{figure}[h]
	\centering
	\includegraphics[width = 0.8\linewidth]{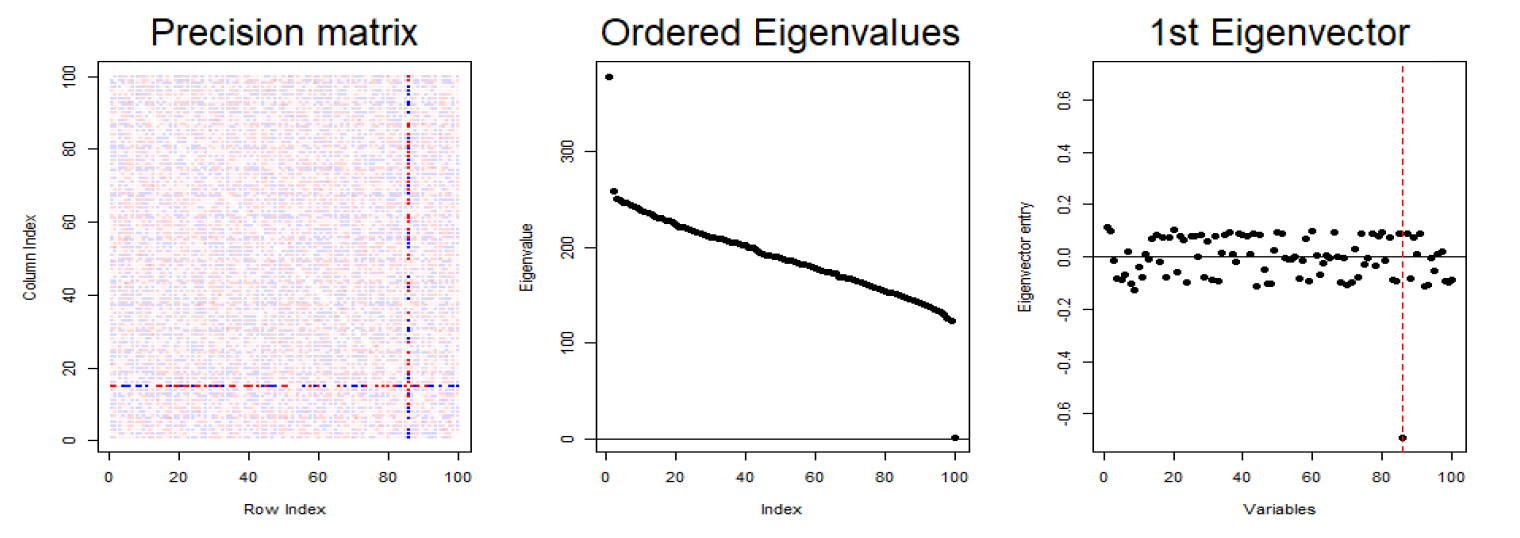}
	\caption{Plots for the illustrative example $\Theta^{(100)}_{2}$. Left panel: visualization of the precision matrix $\Theta^{(100)}_{2}$. The $85$-th variable has large weighted connections to the other variables. Center panel: plot of the ordered eigenvalues of $\Theta^{(100)}_{2}$. The first eigenvalue is visibly separated from the rest. Right panel: coordinates of the first eigenvector of $\Theta^{(100)}_{2}$. The red line highlights the $ 85 $-th coordinate, which has a large magnitude compared to the rest.}
	\label{fig:quick_example2}
\end{figure}

We first consider two examples to illustrate the connection between the presence of highly connected variables in $ \Theta^{(p)} $ and the corresponding spectral decomposition. By these examples, we highlight that if the $ k $-th variable is strongly connected to other variables, then the first eigenvector of $ \Theta^{(p)} $ can detect this variable. These examples represent two possible scenarios of hubs. For the first precision matrix $ \Theta^{(100)}_{1} $, we consider a hub variable $ X_{25} $ with a high degree of connection in the adjacency matrix $ \left[\ind{\Theta^{(100)}_{1,kl} \ne 0}\right]_{1 \le k,l \le 100} $. For the second precision matrix $ \Theta^{(100)}_{2} $, all variables have equivalent degrees in the adjacency matrix $ \left[\ind{\Theta^{(100)}_{2,kl} \ne 0}\right]_{1 \le k,l \le 100} $, but the weight of the connections of the hub variable $ X_{85} $ is significantly larger than the rest. We visualize the matrices $\Theta^{(100)}_{1}$ and $\Theta^{(100)}_{2}$, as well as their eigenvalues and first eigenvectors in Figures \ref{fig:quick_example} and \ref{fig:quick_example2}, respectively. A complete description of the generation of  $\Theta^{(100)}_{1}$ and $\Theta^{(100)}_{2}$  is provided in the Supplementary Materials.

For both examples, the first eigenvalue is visually separated from the rest, and the hub coordinates of the first eigenvectors have much larger magnitudes than the rest. This suggests that the entries of the leading eigenvector of the precision matrix can provide useful information for detecting a single hub variable. We show that the relationship between the presence of hubs and the eigenvalues and eigenvectors of a precision matrix can be extended to the detection of multiple hubs in Sections \ref{subsec:hubs_pdm} and \ref{subsec:hubs_eigenvalues}.

\subsection{Weighted Degree of Connection and Hubs}\label{subsec:hubs_pdm}

We consider a continuous notion of connectivity by interpreting the precision matrix $ \Theta^{(p)} $ as a weighted adjacency matrix of the graphical model $ \cG $. Let $\Theta^{(p)}_{kl}$   denote the $(k,l)$ entry in the matrix $\Theta^{(p)}$. We define the weighted degree of connection for the $ k $-th variable as
\begin{equation}\label{eq:alpha}
	\alpha_{k}^{(p)} := \norm{\Theta^{(p)}_{\cdot k} }{2}^2 = \sum_{l=1}^{p}\big( \Theta^{(p)}_{lk} \big)^{2},
\end{equation}
which is also the $(k,k)$ entry of the squared precision matrix $\big( \Theta^{(Hp)} \big)^{2}$.
This criterion can be interpreted as a convex relaxation of the degree on the discrete adjacency matrix $ \left[\ind{\Theta^{(p)}_{lk} \ne 0}\right]_{k,l \in \cP} $. Specifically, the degree of the $ k $-th variable given by $ \| \Theta^{(p)}_{\cdot k} \|_{0}:= \sum_{l=1}^{p}\ind{\Theta^{(p)}_{lk} \ne 0} $ is relaxed to $ \sum_{l=1}^{p}\big( \Theta_{lk}^{(p)} \big)^{2} $. 
A similar criterion using the $\ell_1$-norm was considered by \citet{tan2014learning} and \citet{McGillivray_2020} in their penalties. 

Another motivation for the criterion \eqref{eq:alpha} is to measure the influence of the $ k $-th variable $ X_{k} $ towards the remaining variables $ \bX_{\cP\backslash \{k\}} $. To be specific, let $ \bX \sim N_{p}(\bmu,\Sigma^{(p)}) $, and $ \bbI(X_{k};\bX_{\cP \backslash \{k\}}) $ be the mutual information between $ X_{k} $ and $ \bX_{\cP \backslash \{k\}} $. Then, under some mild conditions on $ \Sigma^{(p)} $, we can show that $ \exp\{2\cdot\bbI(X_{k};\bX_{\cP\backslash \{k\}})\} $ is asymptotically equivalent to $ \alpha^{(p)}_{k}$, \textit{i.e.} $ \exp\{2\cdot\bbI(X_{k};\bX_{\cP\backslash \{k\}})\} \propto \alpha^{(p)}_{k}$ as $ p \to \infty $. Therefore, the criterion \eqref{eq:alpha} can be interpreted as a quantification of the influence of the $ k $-th node over the remaining variables. In particular, a node $k\in\cP$ with large $ \alpha^{(p)}_{k} $ is highly influential. More details on the connection between the weighted degree of connectivity and the mutual information can be found in the Supplementary Materials.

Based on the connectivity criterion \eqref{eq:alpha}, we define a set of {\it hubs} as variables with higher weighted degrees of connection compared to the other variables. 

\begin{definition}[Hub Set]\label{def:hubs}
	Let $\Theta^{(p)}\in\pdset{p}$ be a precision matrix with $p$ variables in $\cP$. For constants $c_{p} \le \tau_{p}$, we say that a variable subset $\cH^{(p)}\subset\cP$ is a $(\tau_p,c_p)$-{\it set of hubs}, if
	\begin{equation}\label{eq:hub_defn2}
		\frac{\max_{h\in\cH^{(p)}} \alpha^{(p)}_{h} }{\min_{h\in\cH^{(p)}} \alpha^{(p)}_{h}} \leq c_p, \hspace{5mm} \frac{\min_{h\in\cH^{(p)}} \alpha^{(p)}_{h} }{\max_{k\notin\cH^{(p)}} \alpha^{(p)}_{k}} \geq \tau_p.
	\end{equation}
	Moreover, each $ h \in \cH^{(p)}$ is said to be a {\it hub variable} for $\Theta^{(p)}$, and we refer to $\tau_p$ as the {\it separation rate between hubs and non-hubs}.
\end{definition}	

Under Definition \ref{def:hubs}, a set $\cH^{(p)} \subseteq \cP$ is a set of hubs if it satisfies two properties. First, the weighted degree of connection $ \alpha^{(p)}_{h} $ for $ h \in \cH^{(p)} $ must be greater than the non-hub variables in $\cP \setminus \cH^{(p)}$ by at least a factor of $\tau_p$. Second, we require that all $ \{\alpha^{(p)}_{h}\}_{h\in\cH} $ have the same magnitude aside from a factor of $c_p$. To emphasize the separation between hubs and non-hubs, we assume that $1\leq c_p \ll \tau_p$. 

For a particular $p$ and a precision matrix $\Theta^{(p)}\in\pdset{p}$, there may exist several sets $ \cH^{(p)} $ satisfying the properties of \eqref{eq:hub_defn2}, depending on the values of $(\tau_p,c_p)$. In general, the size $ |\cH^{(p)}| $ may not be bounded as $ p \to \infty $, or the separation $\tau_p$ may be small. We are interested in learning how the presence of a finite set of hub variables with a growing separation rate $\tau_p\to \infty$ influences precision matrices $\Theta^{(p)}\in\pdset{p}$ as $p\to\infty$. To this end, consider a fixed $c > 1$, and $\{\tau_p\}_p$ a sequence such that $\liminf_p \tau_p = \infty$. For each $ p \in \bbN $, let $ \Theta^{(p)} \in \pdset{p}$ be a precision matrix, and $ \cH^{(p)} \subseteq \{1,\ldots, p\}$ a set of $(\tau_p, c)-$hubs with $ \sup_p|\cH^{(p)}| < +\infty$. 

The existence of a finite hub set with an increasing separation rate, \textit{i.e.} requiring that $ \sup_{p \in \bbN}|\cH^{(p)}| < +\infty $ and $\tau_p \to \infty$, can imply some restrictions on the matrices $ \{\Theta^{(p)}\}_p $. In Section \ref{subsec:hubs_eigenvalues}, we further discuss the implications of the presence of hub variables on the spectral decomposition of precision matrices.

\subsection{Hubs and Spectral Decomposition}\label{subsec:hubs_eigenvalues}

Denote by $\lambda_{1}^{(p)}\geq \lambda_{2}^{(p)}\geq \ldots\geq \lambda_{p}^{(p)}> 0 $ the eigenvalues of the precision matrix $\Theta^{(p)}$, and by $\bv_{i}^{(p)} = (v_{i1}^{(p)},\cdots,v_{ip}^{(p)})^{\mt} \in \bbR^{p}$ the eigenvector corresponding to the eigenvalue $\lambda_{i}^{(p)}$. 
In this section, we aim to obtain useful implications of the existence of a hub set on the eigenvalues and eigenvectors. 
We consider the following Assumption \ref{as:raw_eigenval_control} on the eigenvalues of $\Theta^{(p)}$.

\begin{assumption}\label{as:raw_eigenval_control}
	Fix $r < +\infty$.
	There exists a universal constant $c < +\infty$, such that for every $p \ge r$, the precision matrix $\Theta^{(p)}$ contains a $(\tau_p, c)$-set of hubs $ \cH^{(p)} $ with $ |\cH^{(p)}| = r $ as in Definition \ref{def:hubs}.
	We further assume that $\tau_{p} \to +\infty$ and $\big( \lambda_{r+1}^{(p)} \big)^{2}\big/\big\{ \sum_{i=r+1}^{p}\big( \lambda_{i}^{(p)} \big)^{2} / (p-r) \big\} = o(\tau_{p})$ as $p \to +\infty$.
\end{assumption}

In addition to considering a set of $r$ hub variables in Assumption \ref{as:raw_eigenval_control}, 
we further require that after removing the leading $r$ eigen-components, the tail eigenvalues are relatively disperse compared to the hub separation. 
To be specific, the ratio between the $(r+1)$-th squared eigenvalue $(\lambda_{r+1}^{(p)})^2$ and the average of squared tail eigenvalues $\sum_{i=r+1}^p (\lambda_{i}^{(p)})^2/{(p-r)}$ grows slower than $\tau_p$ as $p \to +\infty$. 
As a special case, if $\lambda_{r+1}^{(p)}/\lambda_{p}^{(p)}$  is bounded for all $p$, then $\big( \lambda_{r+1}^{(p)} \big)^{2}\big/\big\{ \sum_{i=r+1}^{p}\big( \lambda_{i}^{(p)} \big)^{2} / (p-r)\big\}$ is also bounded, and hence satisfies $o(\tau_{p})$ as $\tau_{p} \to +\infty$. 
As another special case, if $\tau_{p}/p \to +\infty$, then $\big( \lambda_{r+1}^{(p)} \big)^{2}\big/\big\{ \sum_{i=r+1}^{p}\big( \lambda_{i}^{(p)} \big)^{2}/{(p-r)} \big\} \le p-r = o(\tau_{p})$.
That is, a hub separation rate growing super-linearly in $p$ would also be sufficient for the condition in Assumption \ref{as:raw_eigenval_control}.

Note that Assumption \ref{as:raw_eigenval_control} on the eigenvalues is considerably weaker than the common assumptions for precision matrix estimation in GGM, which typically require that all eigenvalues are in the same scale, namely, $\lambda_{1}^{(p)}/\lambda_{p}^{(p)} \le c$ for some universal constant $c < +\infty$ \citep{ren2015asymptotic,cai2016estimating,McGillivray_2020}.
This excludes the possibility of some diverging eigenvalues. 
In the following Proposition \ref{prop:eigenvalue_separation}, we show that under Assumption \ref{as:raw_eigenval_control}, the existence of a finite hub set indeed entails several diverging eigenvalues in $\Theta^{(p)}$. 
Such a phenomenon further implies that estimating the precision matrix could be challenging due to its ill condition.

\begin{prop}\label{prop:eigenvalue_separation}
	If $\Theta^{(p)}$ satisfies Assumption \ref{as:raw_eigenval_control}, then there exists $1\leq s\leq r$, such that
	\begin{equation}\label{eq:eigenvalue_separation}
		\limsup_{p\to \infty} \frac{\lambda_{1}^{(p)}}{\lambda_{s}^{(p)}}< \infty; \quad \liminf_{p\to \infty} \frac{\lambda_{s}^{(p)}}{\lambda_{s+1}^{(p)}}= \infty.
	\end{equation}
\end{prop}

In the literature, the diverging eigenvalue phenomenon was mainly explored in the context of covariance estimation, principal component analysis and factor analysis \citep{shen2016statistics,wang2017asymptotics,fan2018large,cai2020limiting}. In particular, the eigenvalues for the covariance matrix $\Sigma^{(p)}$ were assumed to satisfy $1/\lambda_{p}^{(p)} \ge \cdots \ge 1/\lambda_{p-s'}^{(p)}  \gg 1/\lambda_{p-s'-1}^{(p)} \ge \cdots \ge 1/\lambda_{1}^{(p)}$, where the spike appears at some tail index $p-s'$, and the magnitude $\lambda_{p-s'-1}^{(p)}/\lambda_{p-s'}^{(p)}$ could diverge to $\infty$. This divergence regime is different from the phenomenon in our Proposition \ref{prop:eigenvalue_separation}. In particular, our spike appears at the leading index $s$ in terms of the precision matrix.

One implication of Proposition \ref{prop:eigenvalue_separation} is that, a rank-$s$ approximation via the spiked components $\widetilde{\Theta}^{(s|p)} := \sum_{i=1}^{s}\lambda_{i}\bv^{(p)}_{i}(\bv^{(p)}_{i})^{\mt} $ could incorporate most information in $ \Theta^{(p)} $, with an approximation error $ {\vertiii{\Theta^{(p)} - \widetilde{\Theta}^{(s|p)} }_{2} \big/ \vertiii{ \Theta^{(p)} }_{2}} = {\lambda_{s+1}^{(p)}/\lambda_{1}^{(p)}} \to 0 $ as $ p \to \infty $. 
We mainly utilize this approximation for hub detection.
The implication of a finite hub set on such a low-rank approximation can be vaguely inferred as follows. We can let $\Theta^* = \pare{(}{ \Theta^{(p)}_{ij}\cdot \ind{i ~ \text{or} ~ j\in\cH^{(p)}} }{)}_{i,j\in\cP}$ be the thresholded precision matrix that keeps the rows and columns associated with the hub variables, whose rank is at most $r$. 
Then the column-wise approximation error $\max_{k \in \cP}\big\|\Theta_{\cdot k}^{(p)} - \Theta_{\cdot k}^{*}\big\|_{2}^{2} \le \max_{k \notin \cH^{(p)}}\alpha_{k} $ would be relatively small compared to $\max_{k \in \cP}\big\| \Theta_{\cdot k}^{(p)} \big\|_{2}^{2} = \max_{k \in \cH^{(p)}}\alpha_{k}$ due to the definition of hub. This suggests a potential $r$-rank approximation.

Next, we can further characterize the hubs in $\Theta^{(p)}$ via the leading $s$ eigenvectors $\bv_1^{(p)},\bv_2^{(p)},\ldots,\bv_s^{(p)}$. To this end, we introduce the influence measure for the $ k $-th variable based on the leading $ s $ eigenvectors as
\begin{equation}\label{eq:omega}
	\omega^{(s|p)}_{k} := \Prr_{kk}^{(s|p)} = \sum_{i=1}^{s} \pare{(}{v_{ik}^{(p)}}{)}^2,
\end{equation} 
where $\Prr^{(s|p)} := \sum_{i=1}^s \bv^{(p)}_i(\bv^{(p)}_i)^\mt \in \bbR^{p\times p}$ is the projection matrix onto the space spanned by the first $ s $ eigenvectors $\bspan\{\bv_{i}^{(p)}\}_{i=1}^{s}$, and $ \Prr_{kk}^{(s|p)} $ is the $ k $-th diagonal element of $ \Prr^{(s|p)} $. It can be shown that $ \omega_{k}^{(s|p)} = \sum_{i=1}^{s}(v_{ik}^{(p)})^{2} \le \sum_{i=1}^{p}(v_{ik}^{(p)})^{2} = 1 $, and $ \sum_{k=1}^{p}\omega_{k}^{(s|p)} =  s $. Since $ s $ is typically bounded, we anticipate in general that $ \omega_{k}^{(s|p)} = O(p^{-1}) $ as $ p \to \infty $.

When an influence measure $ \omega_{k}^{(s|p)} $ is relatively large, {\it e.g.} $ \omega_{k}^{(s|p)} = \Omega(1) $, we consider the $ k $-th variable to be highly influential. 
This motivates the following definition of an influential set analogous to the hub set in Definition \ref{def:hubs}. For $\Theta^{(p)}\in\pdset{p}$ and $s< {p}/{2}$, we say that a set $\cI^{(p)}\subset\cP$ is a $(\tau_p, c_p)-${\it influential set} for $1\leq c_p < \tau_p$ if
\begin{equation}\label{eq:omega_dominant}
	\frac{\max_{k\in \cI^{(p)}}\omega_{k}^{(s|p)}}{\min_{k\in \cI^{(p)}}\omega_{k}^{(s|p)}} \leq c_p, \hspace{5mm} \frac{\min_{k\in \cI^{(p)}}\omega_{k}^{(s|p)}}{\max_{k\notin \cI^{(p)}}\omega_{k}^{(s|p)}} \geq \tau_p.
\end{equation}
Such an influential set shares a similar interpretation as the hub set, but in terms of the eigenvector-based influence measure. 
One advantage of this measure is that, it only involves the leading $ s $ eigenvectors of the precision matrix. 
Here, the number of eigen-components $ s $ is implied by the spike in Proposition \ref{prop:eigenvalue_separation}. 
As a consequence, the influence measures can be estimated from a few eigen-components of the sample covariance matrix, without the need of estimating the precision matrix. 
In contrast, the weighted degree measures are defined from all precision matrix elements as in \eqref{eq:alpha}. Their precise estimation would rely on the complete information of the precision matrix. 

Despite the convenience of identifying an influential set, its relationship with the hub set remains unclear. Theorem \ref{thm:equivalence} below establishes that, under an additional influential signal condition, the hub set of interest is also an influential set. This justifies the sufficiency of influence measures for hub detection.

\begin{thm}\label{thm:equivalence}
	Suppose the precision matrix $\Theta^{(p)}$ satisfies Assumption \ref{as:raw_eigenval_control} with the hub set $\cH^{(p)}$. Let $s$ be the spike index in Proposition \ref{prop:eigenvalue_separation}. 
	We further assume the following influential signal condition:
	there exists a universal constant $c < +\infty$, such that
	\begin{equation}\label{eq:assumption_vectors}
		\left( {\lambda_{s+1}^{(p)} \over \lambda_{1}^{(p)}} \right)^{2} \le c\min_{h\in \cH^{(p)}} \omega^{(s|p)}_{h}.
	\end{equation} 
	
	\noindent Then there exists universal constants $0 < c',c'' < +\infty$, such that $\cH^{(p)}$ is a $( \tau_p, c')$-influential set based on the leading $ s $ eigenvectors. Moreover, we have
	\begin{equation*}
		{\min_{h\in \cH^{(p)}}\omega_{h}^{(s|p)} \over \max_{k\notin \cH^{(p)}}\omega_{k}^{(s|p)}} \ge c''\tau_{p}; \quad \min_{h \in \cH^{(p)}}\omega_{h}^{(s|p)} \ge c''\min\left\{ 1, {\tau_{p} \over p} \right\}.
	\end{equation*}
\end{thm}

Theorem \ref{thm:equivalence} establishes another important implication of the presence of hub, and plays a central role in our proposed methodology.
In particular, it suggests that hub detection can be reduced to identifying an influential set. 
Based on the leading $ s $ eigenvectors of the precision matrix, the influence measures  \eqref{eq:omega} are sufficient for identifying such an influential set. By Theorem \ref{thm:equivalence}, this essentially identifies the hub set of interest. Such a strategy can greatly relieve the challenge of estimating hubs based on limited information.

The influential signal condition \eqref{eq:assumption_vectors} is required by Theorem \ref{thm:equivalence}. 
Here, the left hand side is the error due to the rank-$ s $ approximation. 
This approximation error rate should not be greater than the signal provided by the influence measures $ \min_{k \in \cH^{(p)}}\omega_{k}^{(s|p)} $. Such a requirement is necessary for reducing the problem to the leading $ s $ eigenspaces. One sufficient condition for \eqref{eq:assumption_vectors} is that the approximation error is vanishing sufficiently fast:
\begin{equation}\label{eq:strong_connection}
	\left( {\lambda_{s+1}^{(p)} \over \lambda_{1}^{(p)}} \right)^{2} = o\left( {\tau_{p} \over p} \right).
\end{equation}
The small-error condition \eqref{eq:strong_connection} can be easily met. As a special case, if the hub separation rate is at least linear in $p$, that is, $\tau_{p}= \Omega(p)$, then Assumption \ref{as:raw_eigenval_control} can directly imply \eqref{eq:strong_connection} due to Proposition \ref{prop:eigenvalue_separation}. As another special case, if the tail eigenvalues $\lambda_{s+1}^{(p)},\cdots,\lambda_{p}^{(p)}$ are in the same scale such that $\lambda_{s+1}^{(p)}/\lambda_{p}^{(p)}$ is bounded, then for any $\tau_{p}/\sqrt{p} \to \infty$, the small-error condition \eqref{eq:strong_connection} is satisfied. 
More discussions are provided in the Supplementary Materials.

\section{Methodology}\label{section:methodology}

In the rest of the paper, we will omit the superscript $(p)$ for ease of notation.

Suppose $\Theta\in\bbR^{p\times p}$ is a precision matrix that contains a hub set $\cH\subseteq\cP$ as in Assumption \ref{def:hubs}. Our goal is to identify the hub set $ \cH$ based on data $\bX^{1},\bX^{2}, \cdots, \bX^{n} \stackrel{\rm i.i.d.}{\sim} N_p(\bmu,\Sigma)$. Let $\widehat{\bSigma}_{n}\in\bbR^{p\times p}$ be a general estimator for the covariance matrix $\Sigma$, 
with eigenvalues $\bhat\gamma_{1}\geq \bhat\gamma_{2} \geq \ldots \geq \bhat\gamma_{p}$ 
and eigenvectors $\widehat{\bv}_{p}, \widehat{\bv}_{p-1}, \ldots ,\widehat{\bv}_{1}\in\bbR^{p}$. 
In particular, $\widehat{\bv}_{p-i+1}$ corresponds to $\widehat{\gamma}_{i}$.
Here, we allow the flexibility to consider a wide range of covariance estimators, including the sample covariance, screened estimator and masked estimator.
More discussions on the covariance estimators and their statistical properties are provided in Sections \ref{section:theoretical} and \ref{section:simulations}.

Our theoretical background in Section \ref{section:framework} suggests that, in the presence of a hub set $\cH$, (\textit{i}) there is a small number $ s $, such that the eigenvalues of $\Theta$ satisfy
$ {\lambda_{s}}/{\lambda_{s+1}} \gg {\lambda_{1}}/{\lambda_{s}}$;
 (\textit{ii}) the influence measures $\pare{ \{ }{ \omega^{(s)}_k }{ \} }_{k\in\cP} $ satisfy 
$ \min_{k \in \cH}\omega_{k}^{(s)} \gg \max_{k \notin \cH}\omega_{k}^{(s)} $. 
These two facts provide the key ingredients for our methodology, which consists of two steps. First, we obtain an estimator $\bhat s$ for the number of separated eigenvalues $s$. Then, we use the tail $\bhat s$ eigenvectors $\widehat{\bv}_{1},\cdots,\widehat{\bv}_{\widehat{s}}$ of $\widehat{\bSigma}_{n}$ to calculate the influence measure of each variable. The influence measures are then used to classify hubs from non-hubs. We refer to this procedure as {\it Inverse Principal Components for Hub Detection (IPC-HD)}. A summary of the proposed method is provided in Algorithm \ref{alg:ipc_hd}. In what follows, we describe our methodology in detail.

\begin{singlespace}
	\begin{algorithm}
		\SetAlgoLined
		\KwIn{Covariance matrix estimator $ \widehat{\bSigma}_{n}$, regularizer $\rho$, threshold $\kappa\in(0,1]$.} 
		Compute eigenvalues/vectors $ \{ \bhat\gamma_{i}, \widehat{\bv}_{p-i+1} \}_{i=1}^{p} $ of $ \bhat{\bSigma}_n$.\\
		\If{\texttt{method} = ``\texttt{Over-Estimated $\bhat s$}''}{
			Set $\bhat s = \floor{{p}/{5}}$.\\
		}
		\If{\texttt{method} = ``\texttt{Data-Driven $\bhat s$}''}{
			Compute $\delta_{\rho}(i) = {(\bhat\gamma_{p-i} + {\rho})}\big/{(\bhat\gamma_{p-i+1} + {\rho})}$ for $1\leq i\leq \floor{{p}/{2}}$.\\
			Find maximal indices $i_1\neq i_2$ such that $\delta_{\rho}(i_1) \geq \delta_{\rho}(i_2) \geq \max_{j\neq i_1,i_2}\{\delta_{\rho}(j)\}$.\\
			\eIf{$ \delta_{\rho}(i_{1}) > 1.5\cdot \delta_{\rho}(i_{2}) $}{
				set $ \bhat s = i_{1} $; \\
			}{
				set $\bhat s = \floor{{p}/{5}}$.\\
			} 
		}
		For each $k \in \cP$, set $\bhat\omega_{k} = \sum_{i=1}^{\bhat s} v_{ik}^{2}$.\\
		Set $\bhat\cH := \{k\in\cP \,:\, \bhat\omega_k \geq \kappa\}$.\\
		\KwOut{the hub set $ \bhat\cH $.}
		\caption{Inverse Principal Components for Hub Detection}\label{alg:ipc_hd}
	\end{algorithm}
\end{singlespace}

First, we estimate the number of separating eigenvalues $ s $. For this, we introduce a regularizer parameter $\rho > 0$. We define the $i$-th $\rho$-regularized eigenvalue ratio $\delta_{\rho}(i) := {(\bhat\gamma_{p-i} + \rho)}/{(\bhat\gamma_{p-i+1} + \rho)}$ for any $ 1 \leq i \leq \lfloor {p}/{2} \rfloor $. The eigenvalue ratio $\delta_{\rho}(i)$ serves as an approximation of the population ratio ${\lambda_i}/{\lambda_{i+1}}$ of $\Theta$. Since for precision matrices $\Theta$ with hubs, we expect the large ratio $\delta_{\rho}(s)$ for a small value $s$, it is sufficient to calculate $\delta_{\rho}(i)$ for $1\leq i\leq \lfloor {p}/{2} \rfloor$. 

The use of eigenvalue ratios to detect low-rank structures may be challenging with a low sample size. Despite this difficulty, we establish in Theorem \ref{thm:rate2} that an over-estimation $\bhat s > s$ can still be valid, in the sense that the estimated hub set can cover the truth with high probability. From this, we provide two methods for estimating $s$. First, we consider a \textit{data-driven estimation  of $s$}: if there is an index $1\leq \bhat s\leq \lfloor {p}/{2} \rfloor$ such that $\delta_{\rho}(\bhat s) \geq 1.5\cdot \max_{i\neq \bhat s} \delta_{\rho}(i)$, then $\bhat s$ is the estimated number for $s$. If no such index exists, we set $\bhat s = \lfloor {p}/{5} \rfloor$. Our second proposed estimator is the \textit{over-estimation of $s$}, for which we simply set $\bhat s \equiv \lfloor {p}/{5} \rfloor$. 

After obtaining the estimator $ \widehat{s} $, we compute the sample influence measure for each variable $k\in\cP$ as $\widehat{\omega}_{k} = \sum_{i=1}^{\widehat{s}} \widehat{v}_{ik} ^{2} $. To select the hubs, we choose a threshold $\kappa > 0$, and set all variables with influence measures above the value of $\kappa$ as hubs. We denote by $\widehat{\cH}(\rho, \kappa) = \{ k \in \cP :\, \widehat{\omega}_{k}\geq \kappa\}$ the estimated hub set when using the data-driven estimator $\bhat s$. For the over-estimated $\bhat s \equiv S = \floor{{p}/{5}}$, we denote the estimated hub set as $\widehat{\cH}_{S}(\kappa)$.

\section{Theoretical Properties}\label{section:theoretical}

In this section, we establish the hub detection guarantees for IPC-HD based on various covariance estimators. The corresponding sample size requirements are further discussed. The following assumptions are made in our analysis.

\begin{assumption}\label{as:hub_strength}
	There exists a universal constant $r < +\infty$, such that the matrix $\Theta$ contains a set of $r$ hubs $\cH\subset\cP$ satisfying Assumption \ref{as:raw_eigenval_control} with $ \tau_{p} = \Omega(p^{\beta})$ for some $0< \beta \leq 1$. 
\end{assumption}

\begin{assumption}\label{as:strong_eigenval_control}
	Consider the spike index $s$ in Proposition \ref{prop:eigenvalue_separation}.
	There exists a universal constant $c < +\infty$ such that $\lambda_{1}/\lambda_{s} \le c$ and $\lambda_{s+1}/\lambda_{p} \le c$. 
	Moreover, the influential signal condition \eqref{eq:assumption_vectors} holds.
\end{assumption}

\begin{assumption}\label{as:covest_convergence_rate}
	The covariance estimator $\bhat\bSigma_n$ satisfies that, for every $\Delta \ge 0$ sufficiently small, there exists some $\cE(n,p,\Delta) < +\infty$, such that $\vertiii{\bhat\bSigma_n - \Sigma}_2 \leq \vertiii{\Sigma}_2 \cE(n,p,\Delta)$ with probability at least $1 - \Delta$.
\end{assumption}
	
In Assumption \ref{as:hub_strength}, we focus on a precision matrix with hubs and the separation rate $\tau_p = \Omega(p^\beta)$. Here, the parameter $\beta \in (0,1]$ specifies the signal strength of hubs. 
In particular, $\beta > 0$ suggests that the separation is strictly increasing in $p$, so that the hubs are distinguished from the non-hubs asymptotically. 
A greater value of $\beta$ corresponds to a stronger separation between hubs and non-hubs.
If $\beta = 1$, then it corresponds to the strongest separation and the tightest result obtained in our Theorem \ref{thm:rate2} below.
 
Assumption \ref{as:strong_eigenval_control} highlights the requirements on the eigenvalues of precision matrix $\Theta$. 
Specifically, our Proposition \ref{prop:eigenvalue_separation} suggests that a set of $r$ hubs entails at most $r$ spiked eigenvalues in $\Theta$. We further assume that after removing the leading spiked eigenvalues, the remaining ones are in the same scale. As a consequence, the eigenvalue ratio $i\mapsto\lambda_{i}/\lambda_{i+1}$ is bounded at every $i\neq s$ and only spikes at $i = s$, resulting in a unique spike index $s$. 
We emphasize that such a spike leads to an ill-conditioned precision matrix, and its estimation could be challenging.
The extension to multiple spikes with heterogeneous divergence regimes could be more challenging \citep{shen2016statistics}, and we would leave it as a future research direction.

Assumption \ref{as:covest_convergence_rate} plays a central role in the hub detection consistency of IPC-HD. It assumes a generic concentration condition on covariance estimation in terms of spectral norm.
This generic condition will be further specialized to various covariance estimators, including
the sample covariance with $\cE(n,p,\Delta) \propto \sqrt{\max\{p,\log\Delta^{-1}\}/n}$ \citep{koltchinskii2017concentration}, 
and the structured estimators with  $\cE(n,p,\Delta) \propto \sqrt{\max\{\mathsf{complexity} \times \log(p), \log \Delta^{-1}\} / n}$ for some complexity measure. Detailed discussions are provided in Sections \ref{subsec:hubest_theory_screened} and \ref{subsec:hubest_theory_mask}.

Based on Assumptions \ref{as:hub_strength}, \ref{as:strong_eigenval_control}, \ref{as:covest_convergence_rate}, we establish the high-probability guarantee of hub detection for IPC-HD in the following Theorem \ref{thm:rate2}.

\begin{thm}\label{thm:rate2}
	Assume that the precision matrix $\Theta$ satisfies Assumptions \ref{as:hub_strength}, \ref{as:strong_eigenval_control}, 
	and the covariance estimator $\bhat\bSigma_n$ satisfies Assumption \ref{as:covest_convergence_rate}. 
	There exists a universal constant $c < +\infty$, such that for every $\Delta \ge 0$ sufficiently small, $(n,p)$ sufficiently large such that $\cE(n,p,\Delta) \le cp^{-(1-\beta)}$, and some $\kappa,\rho > 0$, we have:
	\begin{enumerate}[label=(\Roman*)]
		\item (Data-Driven $\widehat{s}$) \label{item:thm2_result_s} $\bhat\cH(\rho, \kappa) = \cH$ with probability at least $1 - 2\Delta$;
		\item (Over-Estimated $\widehat{s}$) \label{item:thm2_result_S} $\bhat\cH_{S}(\kappa)\supseteq \cH$ with probability at least $1 - \Delta$.
	\end{enumerate}
\end{thm}

Theorem \ref{thm:rate2} quantifies the required sample size $n$ and dimension $p$ for hub detection with high probability. 
In particular, under Assumption \ref{as:hub_strength} with a hub separation rate $\tau_p= \Omega(p^\beta)$, it requires sufficiently large $n,p$ such that the covariance estimation error $\cE(n,p,\Delta)$ is below a factor of $p^{-(1-\beta)}$. 
As $\beta$ increases from $0$ to $1$, that is, the hub signal becomes stronger, such a requirement is less demanding. With the strongest hub signal such that $\beta = 1$, it only requires that the covariance estimation error $\cE(n,p,\Delta)$ goes below a universal constant.
To our best knowledge, Theorem \ref{thm:rate2} is the first hub recovery result from which stronger hub signal implies a faster convergence result. 

Two versions of IPC-HD in Algorithm \ref{alg:ipc_hd} are justified in Theorem \ref{thm:rate2} with the high-probability statements. 
In Part \ref{item:thm2_result_s}, 
we show that based on our data-driven $\widehat{s}$, the estimated hub set $\bhat\cH(\rho,\kappa)$ is a perfect recovery of the true hub set $\cH$. 
In Part \ref{item:thm2_result_S}, 
we further show that based on the over-estimated $\widehat{s} = S$, the resulting hub set estimator $\bhat\cH_{S}(\kappa)$ can still contain the true hub set $\cH$. 
The second statement confirms the validity of over-estimating $s$ as a reliable method for hub detection, even though false hub variables may be detected.
It could be applicable when shortlisting hub candidates is helpful, while identifying the true spike index $s$ remains challenging.
In the Supplementary Materials, additional discussions and simulation studies are provided for over-estimating $s$ and its impacts on IPC-HD.

Our Theorem \ref{thm:rate2} allows the flexibility to be applied to various covariance estimators. 
As a canonical application, the sample covariance $\bhat\Sbb_{n}$ satisfies the concentration condition in Assumption \ref{as:covest_convergence_rate} with an estimation error $\cE(n,p,\Delta) 
\propto \max\left\{ \sqrt{p/n}, p/n, \sqrt{(\log \Delta^{-1}) / n}, {(\log \Delta^{-1}) / n} \right\}$ \citep{koltchinskii2017concentration}. 
For $n \ge \max\{p,\log \Delta^{-1}\}$, it can be further simplified as $\cE(n,p,\Delta) \propto \sqrt{\max\{p,\log\Delta^{-1}\}/n}$,
and the corollary for hub detection is obtained below.

\begin{cor}[IPC-HD with Sample Covariance]\label{cor:hubest_samplecov}
	Suppose $\Theta$ satisfies Assumptions \ref{as:hub_strength} and \ref{as:strong_eigenval_control}, and consider the sample covariance estimator $\bhat\bSigma_n=\bhat\Sbb_n$. 
	Then there exists a universal constant $c < +\infty$, such that for every $\Delta \ge 0$, $n \ge c\max\{p^{3-2\beta},p^{2-2\beta}\log\Delta^{-1}\}$, and some $\kappa,\rho> 0$, we have: 
	(I) $\bhat\cH(\rho, \kappa) = \cH$ with probability at least $1 - 2\Delta$; and (II) $\bhat\cH_{S}(\kappa) \supseteq \cH$ with probability at least $1 - \Delta$.
\end{cor}

In Corollary \ref{cor:hubest_samplecov} with $\Delta = e^{-p}$, the sample size requirement for successful hub detection is $n = \Omega(p^{3-2\beta})$. In particular, for $\beta = 1$ corresponding to the strongest separation rate $\tau_p = \Omega(p)$ in Assumption \ref{as:hub_strength}, it only requires a sample size $n = \Omega(p)$. This result is driven by the fact that, the sample covariance estimator requires $n = \Omega(p)$ for accurate covariance estimation in terms of the spectral norm \citep{koltchinskii2017concentration}.

To tackle with the high-dimensional cases when $n \ll p$, we make additional assumptions on the covariance structure, and further consider alternative covariance estimators to exploit these structural assumptions. 
In Section \ref{subsec:hubest_theory_screened}, we consider an approximately block-diagonal structure, such that the majority of connections concentrate on a $T$-by-$T$ principal submatrix of $\Theta$ for some $T \le p$. The screened covariance estimator can guarantee consistent hub detection even for $n \propto \log(p)$.
In Section \ref{subsec:hubest_theory_mask}, we consider the availability of covariance masking, such that the masked covariance is close to the truth, and can be estimated with a weak sample size requirement. The thresholded covariance estimator under certain sparsity assumption is a special case, and we can establish its hub detection consistency for $n \propto \log (p)$.
To complement our theoretical results in this section, we provide comprehensive simulations in Section \ref{section:simulations} under the low- and high-dimensional settings to further explore the performance of IPC-HD based on various covariance estimators.

\subsection{IPC-HD on Screened Covariance Matrix}\label{subsec:hubest_theory_screened}

Suppose that there exists a subset $\cT \subseteq\cP$ with $|\cT| = T$, such that after variable reordering, the precision matrix $\Theta$ takes the following form
\begin{equation}\label{eq:ScreeningStructure}
	\Theta = \begin{pmatrix}
		\Theta_{\cT} & \Er^\mt \\ \Er & \Fr
	\end{pmatrix},
\end{equation}
where $\Theta_{\cT} \in\bbR^{T\times T}$ is a signal block, and $\Er\in\bbR^{(p-T)\times T}$ and $\Fr\in\bbR^{(p-T)\times (p-T)}$ are non-signal blocks.
Based on the decomposition \eqref{eq:ScreeningStructure}, we assume that for some $\zeta > 1/2$, we have $\vertiii{ \Er }_2,\vertiii{ \Fr -\bdiag(\Fr)}_2 = O(T^{-\zeta})$;
and in terms of the corresponding variance-covariance, for some $\eta < 1/2$, we have $\min_{k\in\cT}\max_{i\neq k} |\sigma_{ik}|= \Omega(T^{-\eta})$.

Note that our structural assumptions essentially require that both the precision matrix $\Theta$ and covariance matrix $\Sigma$ are approximately block-diagonal, so that they could be block-wise inverse of each other.
The exact block-diagonal special case is $\Er = 0$.
The block-diagonal structures have been widely explored in a variety of settings to improve the estimation of graphical models \citep{witten2011new,mazumder2012exact, danaher2014joint, mohan2014node, tan2014learning}.
In our setting \eqref{eq:ScreeningStructure}, the non-signal blocks are negligible in terms of small $\vertiii{ \Er }_2$ and $\vertiii{ \Fr -\bdiag(\Fr)}_2$. In this way, the corresponding blocks in the covariance matrix could have negligible entries, that is, $\max_{i \notin \cT\, \text{or}\, k \notin \cT}|\sigma_{ik}|$ could be small. In addition, the signal block is recovered via a large $\min_{k\in\cT}\max_{i\neq k} |\sigma_{ik}|$.
Similar approximate block-diagonal and minimal signal structures were also studied in \cite{luo2014sure, mohan2014node, liang2015equivalent, kuang2017screening, liang2022markov}.
Under such structures, the variable screening based on sample covariance or correlation can greatly reduce the effective dimension before further analysis, and help the estimation of high-dimensional graphical models accurately even with a low sample size. In genetic or neurological applications, it was also reported that most of the graphical structures of practical interest may be concentrated on a small subset of relevant features \citep{jia2017learning, qiu2020estimating}. This confirms the relevance of block-diagonality in real world scenarios.

Our covariance screening and IPC-HD would be proceeded in the following two steps. 
Let $\bhat s_{ij}$ denote the $(i,j)$ entry of $\bhat\Sbb_n$. Fix $T \le n$ as a desired size of screening.
In the first step, for each variable $i \in \cP$, we compute its largest absolute covariance with the remaining variables $\max_{j \neq i} |\widehat{s}_{ij}|$.
After sorting these measures across all variables, we obtain the variable subset $\widehat{\cT} \subseteq\cP$ from the top $T$ largest measures. 
In the second step, we apply our IPC-HD method to the screened sample covariance $\bhat\Sbb_{n}(\bhat\cT) = [\widehat{s}_{ij}]_{i,j \in \widehat{\cT}}$ as a $T$-by-$T$ principal submatrix of $\bhat\Sbb_{n}$ restricted to the screened subset $\bhat\cT$. 

We denote by $\bhat\cH(T, \rho, \kappa)$ and $\bhat\cH_S(T,\kappa)$ as the two versions of the post-screening IPC-HD hub set estimators in Algorithm \ref{alg:ipc_hd}. 
Under the structural assumptions based on \eqref{eq:ScreeningStructure} and a bounded diagonal assumption, we develop the post-screening hub detection guarantee in the following Corollary \ref{cor:hubest_screenedcov}. A formal description of the assumptions for Corollary \ref{cor:hubest_screenedcov} and its proof can be found in our Supplementary Materials.

\begin{cor}[IPC-HD with Covariance Screening]\label{cor:hubest_screenedcov}
	Suppose $\Theta$ takes the form in \eqref{eq:ScreeningStructure} and satisfies the assumptions outlined in the Supplementary Material. 
	Then there exists a universal constant $c < +\infty$, such that for every $0 \le \Delta \le \min\{ p^{-1}, e^{-T} \}$, $n\geq c \max\{ T^{2\eta}\log(p), T^{3-2\beta}\}$ and some $\kappa, \rho > 0$, we have: (I) $ \bhat\cH(T, \rho, \kappa) = \cH$ with probability at least $1 - 2\Delta$; and (II) $ \cH \subseteq \bhat\cH_{S}(T, \kappa)$ with probability at least $1- \Delta$.
\end{cor}

Note that the required sample size in Corollary \ref{cor:hubest_screenedcov} is based on the trade-off between the effective dimension $T$ and $\log(p)$. In practice, we typically consider $T \lesssim \log(p)$, which assumes that the signal block in \eqref{eq:ScreeningStructure} has a size of order $\log(p)$. 
If $\beta = 1$, which corresponds to the strongest hub separation rate in Assumption \ref{as:hub_strength}, then the required sample size for hub detection consistency becomes $n=\Omega(T^{2\eta}\log(p))$.
It also depends on the exponent $\eta < 1/2$ for the signal covariance strength.
In the ideal case that $\eta = 0$, a sample size $n \propto \log(p)$ would be sufficient.

\subsection{IPC-HD on Masked Covariance Matrix}
\label{subsec:hubest_theory_mask}

In this section, we consider a more general masked estimation framework in high dimension that goes beyond the approximate block-diagonality in \eqref{eq:ScreeningStructure}. 
Consider $\Mr$ as a $p$-by-$p$ \textit{mask matrix}. Instead of estimating $\Sigma$ directly, we consider to estimate the masked covariance matrix $\Mr \odot \Sigma$ as an approximation of $\Sigma$, where $\odot$ denotes the matrix entrywise multiplication.
This is motivated from the \textit{masked covariance estimator} $\bhat\bSigma_n  = \Mr \odot \bhat\Sbb_{n}$, where $\bhat\Sbb_{n}$ is the sample variance \citep{levina2012partial,chen2012masked}. Suppose the $(i,j)$ entry of $\Mr$ is $m_{ij}$, and consider the matrix 1-2 norm $\vertiii{\Mr}_{1,2}^{2} := \max_{j}\sum_{i}m_{ij}^{2}$.
It can be shown that such a masked estimator could satisfy Assumption \ref{as:covest_convergence_rate} with $\cE(n,p,\Delta) \propto \vertiii{\Mr}_{1,2}\sqrt{\max\{\log(p),\log(\Delta^{-1})\}/n}$. The complexity measure $\vertiii{\Mr}_{1,2}^{2}$ is typically small under suitable structural assumptions 
including sparsity \citep{el2008operator,bickel2008covariance},
banding \citep{bickel2008regularized} and tapering \citep{furrer2007estimation,cai2010optimal}.

We first focus on the setting with a known and fixed mask matrix as in \citet{levina2012partial}. 
To establish Assumption \ref{as:covest_convergence_rate}, we first note the following bias-variance decomposition
\[ \vertiii{\bhat\bSigma_n - \Sigma}_{2} \le \underbrace{\vertiii{\Mr\odot\bhat\Sbb_n - \Mr\odot\Sigma}_{2}}_{\text{variance}} + \underbrace{\vertiii{\Mr\odot\Sigma - \Sigma}_{2}}_{\text{bias}}. \]
In the following Assumption \ref{as:mask_bias},
we further assume that the mask matrix $\Mr$ incurs a bias term comparable with the variance term. 

\begin{assumption}\label{as:mask_bias}
	There exists a universal constant $c < +\infty$ such that $\vertiii{\Mr\odot\Sigma - \Sigma}_{2} \le c\vertiii{\Sigma}_{2}\vertiii{\Mr}_{1,2}\sqrt{\log(p)/n}$.
\end{assumption}

One special case of Assumption \ref{as:mask_bias} is that $\Mr$ indicates the non-zero entries of $\Sigma$, that is, $m_{ij} = \ind{\sigma_{ij} > 0}$, so that the bias term is exactly zero. 
More general examples could be referred to \cite{chen2012masked,chen2012masked.old} for blocked, banded and tapered structures. 
In these special cases, the bias term is generally not zero, and the complexity measure $\|\Mr\|_{1,2}^{2}$ is typically bounded. Then Assumption \ref{as:mask_bias} suggests that the approximation error in spectral norm is at most the order of $\sqrt{\log(p)/n}$.

In most applications, the mask matrix is unknown and needs to be estimated from data. This can be done based on a separate independent sample, so that conditional on the separate sample, the mask can be considered as a fixed matrix.
In this case, Assumption \ref{as:mask_bias} also quantifies the requirement for the  mask estimation error.

Based on Assumption \ref{as:mask_bias}, we could further justify Assumption \ref{as:covest_convergence_rate} with $\cE(n,p,\Delta) \propto \vertiii{\Mr}_{1,2}\sqrt{\max\{\log(p), \log(\Delta^{-1})\}/n}$. This can conclude the corollary for hub detection below analogously to the sample covariance estimator.

\begin{cor}[IPC-HD with Masked Covariance Estimator]\label{cor:hubest_mask}
	Suppose $\Theta$ satisfies Assumptions \ref{as:hub_strength} and \ref{as:strong_eigenval_control}, 
	and consider the masked covariance estimator $\bhat\bSigma_n \equiv  \Mr \odot \bhat\Sbb_{n}$ with a mask matrix $\Mr \in \bbR^{p\times p}$ satisfying Assumption \ref{as:mask_bias}. Then there exists a universal constant $c < +\infty$, such that for every $\Delta \ge 0$,
	$n > c\vertiii{\Mr}_{1,2}^{2}p^{2(1-\beta)}\max\{\log(p), \log\Delta^{-1}\}$, and some $\kappa,\rho> 0$, we have: 
	(I) $\bhat\cH(\rho, \kappa) = \cH$ with probability at least $1 - 2\Delta$; and (II) $\bhat\cH_{S}(\kappa) \supseteq \cH$ with probability at least $1 - \Delta$.
\end{cor}

Suppose $\Delta = p^{-\| \Mr \|_{1,2}^{2}}$.
Compared with Corollary \ref{cor:hubest_samplecov}, the sample size requirement in Corollary \ref{cor:hubest_mask} reduces to $n = \Omega(\vertiii{\Mr}_{1,2}^{2}p^{2(1-\beta)}\log(p))$. If $\beta = 1$ in Assumption \ref{as:hub_strength} corresponding to the strongest eigenvalue separation rate, then it only requires a sample size of order $\vertiii{\Mr}_{1,2}^{2}\log(p)$ for consistent hub detection. With a small complexity measure $\vertiii{\Mr}_{1,2}^{2}$, for example, $\vertiii{\Mr}_{1,2}^{2}$ is bounded, then this could allow $n \propto \log(p)$. 

For the rest of this section, we specifically assume that the covariance matrix $\Sigma$ is sparse, in the sense that the maximal number of column-wise non-zero element is at most a universal constant. 
\begin{assumption}\label{as:sparse}
	Let $\Mr$ be a mask matrix with the $(i,j)$ entry $m_{ij} = \ind{\sigma_{ij} > 0}$. Assume that $\| \Mr \|_{1,2}^{2} \le \varsigma$ for some universal constant $\varsigma < +\infty$. That is, the maximal number of non-zero elements per row in $\Sigma$ is at most $\varsigma$.
\end{assumption}
We consider the estimated thresholding mask $\widehat{\Mr}$ to estimate the location of sparsity, where its $(i,j)$ entry takes the form $\widehat{m}_{ij} = \ind{\widehat{s}_{ij} \ge \xi\sqrt{(\log(p)) / n}}$, and $\widehat{s}_{ij}$ is the $(i,j)$ entry of the sample covariance matrix $\bhat\Sbb_{n}$, $\xi \ge 0$ is a tuning parameter.
In this case, the corresponding masked covariance estimator $\widehat{\Mr}\odot\bhat\Sbb_{n}$ is the same as the thresholded covariance estimator \citep{el2008operator,bickel2008covariance}. 
The hub detection guarantee is established below.

\begin{cor}[IPC-HD with Covariance Thresholding]\label{cor:hubest_threshold}
	Suppose the precision matrix $\Theta$ satisfies Assumptions \ref{as:hub_strength}, \ref{as:strong_eigenval_control} and \ref{as:sparse}, 
	and consider the thresholded covariance estimator $\bhat\bSigma_n =  \widehat{\Mr} \odot \bhat\Sbb_{n}$. 
	Then there exists a universal constant $c < +\infty$, such that for every $n > c\varsigma^{2} p^{2(1-\beta)}\max\{\log(p), \log\Delta^{-1}\}$, and some $\kappa,\rho,\xi> 0$, we have: 
	(I) $\bhat\cH(\rho, \kappa) = \cH$ with probability at least $1 - 2\Delta$; and (II) $\bhat\cH_{S}(\kappa) \supseteq \cH$ with probability at least $1 - \Delta$.
\end{cor}

Suppose $\Delta = p^{-\varsigma}$. Following the same discussion as in Corollary \ref{cor:hubest_mask}, the sample size requirement in this case is $n = \Omega(\varsigma^{2}p^{2(1-\beta)}\log(p))$. If $\beta = 1$ in Assumption \ref{as:hub_strength} corresponding to the strongest eigenvalue separation, then it only requires a sample size $n \propto \varsigma^{2}\log(p)$ for consistent hub detection.

\section{Simulation Study} \label{section:simulations}

To validate the effectiveness of the proposed IPC-HD, we perform several simulation studies. We compare our IPC-HD method to other existing methods for hub detection in the literature in high-dimensional settings. The results are summarized and discussed in Section \ref{subsec:sim_comparison}. Additional simulations assessing the sensitivity of the hub detection method to the choice of $\bhat s$ and the performance of IPC-HD combined with variable screening in high dimensional datasets can be found in the Supplementary Materials.

\subsection{Data Generating Process}\label{subsec:data_generation}

To simulate data from a Gaussian distribution with hubs, we start by generating a precision matrix with hubs $\Theta$. Given $n$, we generate $\bX^1,\dots,\bX^n\stackrel{\rm i.i.d.}{\sim} N(\bzero, \Theta^{-1})$. The positive definite precision matrix $\Theta$ is generated as follows. First, we generate a symmetric matrix $\widetilde{\Theta}$ in the form of
\begin{equation*}
	\widetilde{\Theta} = \begin{pmatrix} 
		\widetilde{\Theta}_{\cT} & \Er^{\mt}_{T\times(p -T)}\\ \Er_{(p-T)\times T} & \Fr_{(p-T)\times (p-T)} 
	\end{pmatrix},
\end{equation*} 
where $ T \leq p $. 
To generate the signal block $\widetilde{\Theta}_{\cT}$, we first specify the {\it number of hubs} $r$, the {\it hub connection probability} $p_{\rm H}$, and the {\it non-hub (NH) connection probability} $p_{\rm NH}$, where $0\leq p_{\rm NH} < p_{\rm H} \leq 1$.
We then choose the set of hubs $\cH\subset\cT$ at random, and generate a $T$-by-$T$ adjacency matrix $\Ar$ with independent entries, such that the hub variables $h\in\cH$ are connected to any other variable with probability $p_{\rm H}$, and any pairs of non-hubs are connected with probability $p_{\rm NH}$. 
We further generate a $T$-by-$T$ symmetric matrix $\Ur$ such that $\Ur_{ij} = \Ur_{ji}\sim \Unif([-5,-4]\cup [4,5])$ independently.
The signal block is finally obtained as $\widetilde{\Theta}_{\cT} = \Srr \odot \Ur$, which is a symmetric matrix with zero diagonals, continuous-value off-diagonals, and the sparsity pattern indicated by $\Ar$. 
The non-signal blocks $\Er_{(p-T)\times T}$ and $\Fr_{(p-T)\times (p-T)}$, whose dimensions are in the subscripts, are obtained in the same manner, but with sparser patterns. 
In particular, the \textit{non-signal (NS) connection probability} $p_{\rm NS}$ is specified as a smaller number than $p_{\rm NH}$. The diagonals of $\Fr_{(p-T)\times (p-T)}$ are left zero.

Next, we find the smallest eigenvalue of $\widetilde{\Theta}$ as $\widetilde{\lambda}_{p}$, which is negative, and fix some $\Delta > 0$. 
Then we let the diagonals of $\Theta$ be the same as $\Delta + (- \widetilde{\lambda}_{p})$, and the off-diagonals of $ \Theta $ be the same as $\widetilde{\Theta}$. In this way, the smallest eigenvalue of $\Theta$ is $\Delta > 0$, so that $\Theta$ is positive definite.
The resulting precision matrix $\Theta$ contains $r$ hub variables, $T - r$ variables with low connections to each other but high connections to the hub variables, and $p-T$ variables with low connections to each other and other variables. 
To re-scale the generated matrices, we obtain the correlation matrix $\Rr$ from the covariance $\Theta^{-1}$, and the inverse correlation matrix as $\Xi = \Rr^{-1}$.

\subsection{Comparison with Existing Methods}\label{subsec:sim_comparison}

We compare the following methods for hub detection. First, we consider our proposed IPC-HD method. To explore the performance of our IPC-HD with different choices of the initial matrix estimator,  we include in our simulations the IPC-HD applied to thresholded and screened correlation matrix estimators. In addition to our proposal, we include several existing methods for graphical model estimation in the literature. We consider the graphical LASSO (GLASSO) \citep{friedman2008sparse}, hub GLASSO (HGL) \citep{tan2014learning}, and hub-weighted GLASSO (HWGL) \citep{McGillivray_2020}. In addition, we also consider the {\it raw correlation inversion}, which consists of computing the inverse of the sample correlation matrix $\bhat\bXi^{\text{raw}} = \bhat \Rr_n^{-1}$. When $p> n$, for the raw inverse correlation method we first apply variable screening on the sample correlation matrix.

For each method, we select the hub set based on a measurement of connectivity. For the GLASSO, HGL and HWGL, we use the discrete degree of connectivity. For the raw inverse correlation method, we use the weighted degree of connectivity, as in \eqref{eq:alpha}. For the IPC-HD, we consider the influence measures $(\bhat\omega_{j})_{j\in\cP}$. For each method, once we obtain the measurements of connectivity, we set the estimated set of hubs as the variables with connectivity greater than 2 standard deviations above the average connectivity level.

The graphical models are simulated according to the data generating method described in Section \ref{subsec:data_generation}. We consider the total dimensions $p\in \{ 100, 200, 500\}$, submatrix size $T \in \{0.25p \, , \, 0.5p, p\}$, sample sizes $n\in \{0.25p, 0.50p, 0.75p, p\}$, and $\Delta \in \{2,5\}$. We simulate both weak hubs ($p_{\rm H} = 0.4$, $p_{\rm NH} = 0.05$), and strong hubs ($p_{\rm H} = 0.8$, $p_{\rm NH}= 0.05$). For all simulations, we set $p_{\rm NS} = 0.005$. We report the simulation outcomes for $p =200, 500$ and $\Delta = 5$ in Section \ref{subsec:sim_results}. The remaining results can be found in the Supplementary Materials.

To fit the GLASSO and HWGL, we select the tuning parameters from a fine grid of values, by minimizing the Bayesian information criteria (BIC) \citep{gao2012tuning}. For the HGL, three tuning parameters are selected from a three-dimensional grid by minimizing the modified BIC proposed for the HGL method \citep{tan2014learning}.

We consider three measurements of performance. First, we measure the true positive and false positive rates of hub detection, given by, 
$\TPR(\bhat\cH) =  {\pare{|}{\cH\cap \bhat\cH}{|} }/{\pare{|}{\cH}{|} }$ and $\FPR(\bhat\cH) = {\pare{|}{\cH^{c}\cap \bhat\cH }{|} }/{ \pare{|}{\cH^c}{|} }$. Both $\TPR$ and $\FPR$ are between 0 and 1. A large value of $\TPR$ is preferred, since it indicates most hubs are correctly identified. On the other hand, a small value of $\FPR$ is preferred, since it suggests most non-hubs are classified as non-hubs. Finally, we measure the computational time required to complete the estimation problem.

Since, for any method, the set of estimated hubs consists of the variables with measurements of connectivity 2 standard deviations above the mean, the FPR is controlled below $5\%$ for all methods, and in most cases is significantly below $5\%$. From this, in Section \ref{subsec:sim_results} we only report the $\TPR$. Visualization of the $\FPR$ and computational time for our simulations can be found in the Supplementary Materials.

\subsection{Simulation Results}\label{subsec:sim_results}

\begin{figure}
	\centering
	\includegraphics*[width = 0.8\linewidth]{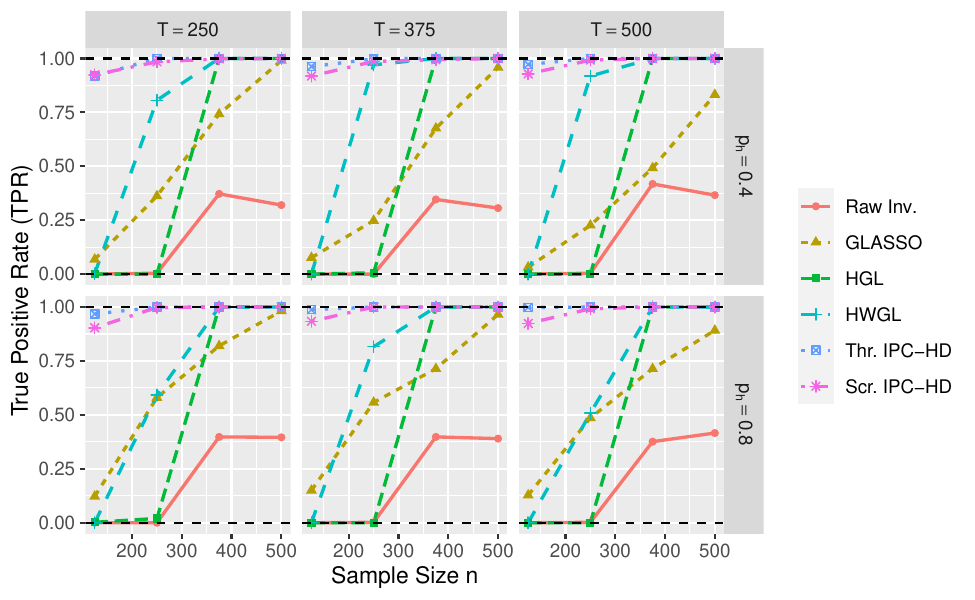}
	
	\caption{Comparison of the simulated true positive rate of hub detection (TPR) across methods for $p = 500$ and $\Delta = 5$. We observe near-perfect recovery for both the thresholded and screened IPC-HD even for our lowest choices of sample size $n$. Our methods are consistent, even when the block-diagonal assumption $T< 500$ is not satisfied. This confirms the reliability of our IPC-HD method for the recovery of hubs in high dimensional GGMs.}\label{fig:tfig_p500d5}
\end{figure}

Our numerical comparison of the TPR for $p = 500$ and $\Delta = 5$ can be found in Figure \ref{fig:tfig_p500d5}. As observed in Figure \ref{fig:tfig_p500d5}, both the screened and thresholded IPC-HD obtain perfect recovery of hubs for all simulations with $n\geq 250$, and have near-perfect recovery for our lowest sample size of $n = 125$. Our results confirm the consistency of our proposed IPC-HD method for recovering hubs in high-dimensional data scenarios, even when the block diagonal assumption \eqref{eq:ScreeningStructure} is not satisfied, \textit{i.e.} $T = 500$.

\begin{figure}
	\centering
	\includegraphics*[width = 0.8\linewidth]{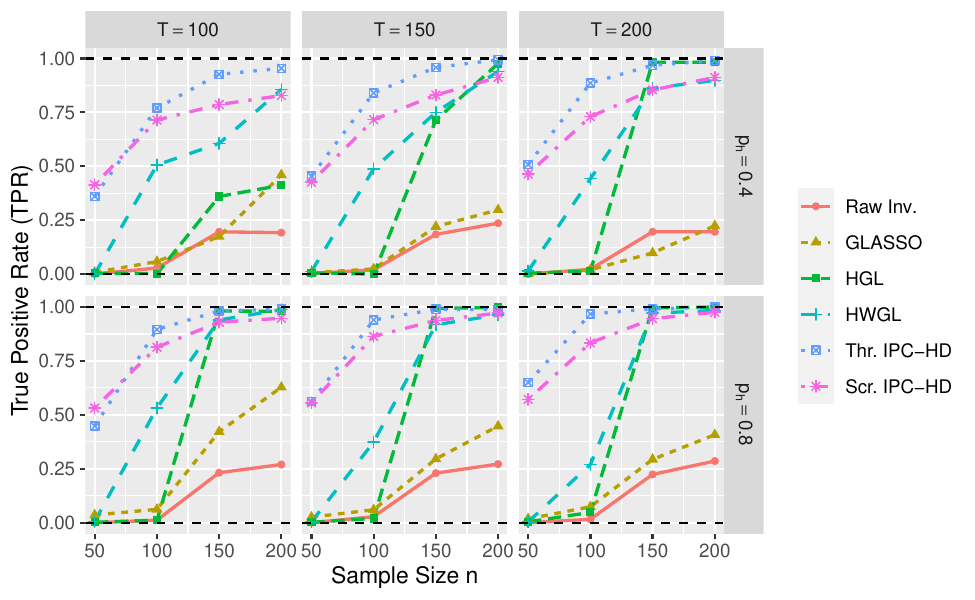}
	
	\caption{Comparison of the simulated true positive rate of hub detection (TPR) across methods for $p = 200$ and $\Delta = 5$. Our thresholded IPC-HD outperforms all other methods in terms of TPR, regardless of the hub strength or size of the block structure $T$. Furthermore, our screened IPC-HD outperforms all other methods in low sample sizes, except in the case that breaks the block-diagonal assumption $T < 200$ is not satisfied.}\label{fig:tfig_p200d5}
\end{figure} 

To explore our guarantees in more challenging scenarios, we provide our numerical results for $p = 200$ and $\Delta = 5$ in Figure \ref{fig:tfig_p200d5}. In Figure \ref{fig:tfig_p200d5}, our thresholded IPC-HD method obtains the highest TPR for all simulation scenarios, regardless of the hub connectivity level $p_h \in \{0.4,0.8\}$ or the size of the submatrix $\Theta^{*}_{\cT}$. In particular, our thresholded IPC-HD obtains favorable results, even when the block structure \eqref{eq:ScreeningStructure} is not satisfied, \textit{i.e.} the size of the submatrix $\Theta_{\cT}$ is $T = p$. Furthermore, our screened IPC-HD outperforms all other methods in terms of hub recovery for the low sample sizes of $n = 50, 100$. When the block structure \eqref{eq:ScreeningStructure} is not satisfied, the screened IPC-HD remains competitive when compared to the HGL and HWGL for the sample sizes $n = 150, 200$.

To provide further insight on other measures of numerical performance such as the false positive rate of hub detection (FPR), computational time, and hub versus non-hub degree comparisons, we include additional visualizations and discussions of these measurements in our Supplementary Materials.

\section{Real Data Application}\label{section:real_data}

We analyze a real dataset containing the gene expression levels of 551 prostate cancer patients\footnote{Dataset available at \url{https://portal.gdc.cancer.gov/}.}. For each patient, more than 60,000 measurements of gene expression levels are available. The goal of our analysis is to find a set of hub genes with a potential high influence over the rest. For each gene, we apply a shifted log-transformation to make the resulting data close to a Gaussian distribution \citep{Feng_2016}. Then, we select the top 200 genes with the largest variances \citep{tan2014learning}. To eliminate the potential effect of clinical variables in our analysis, we remove from all genes the effects of age, race, initial diagnosis, vital status and whether the patient received radiation or pharmaceutical treatment before the sample collection. 

\begin{figure}[h]
	\centering
	\includegraphics[width = 0.7\linewidth]{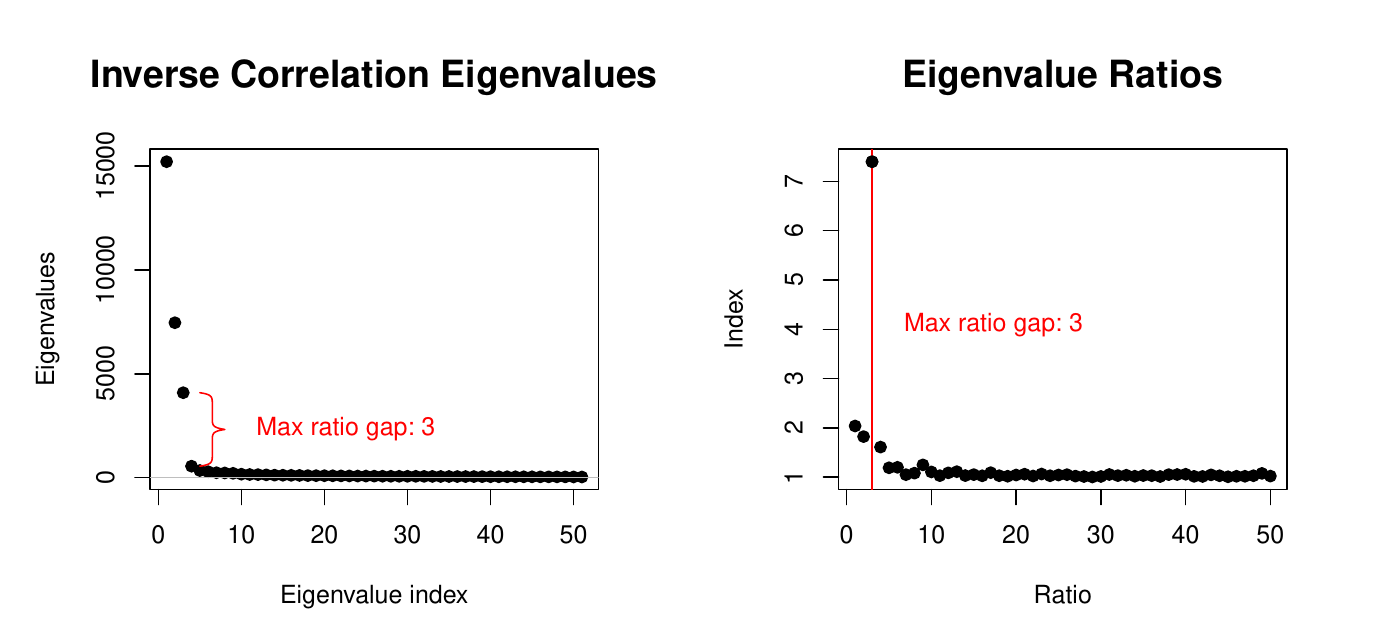}
	\caption{Left panel: plot of the ordered eigenvalues $\{\lambda_i\}_{i=1}^{50}$ of the empirical inverse correlation matrix. We observe three separating eigenvalues. Right panel: plot of the first 50 top eigenvalue ratios $\{{\lambda_{i}}/{\lambda_{i+1}}\}_{i=1}^{50}$  of the inverse correlation matrix. We confirm the presence of an eigenvalue separation at $\bhat s = 3$.}
	\label{fig:prostate_eigenvalue_gaps}
\end{figure}

To estimate the hub genes in this dataset, we calculate the empirical correlation matrix $\bhat \Rr_{200}\in \pdset{200}$, and the eigenvalues of the inverse correlation matrix $\bhat\bXi_{200}=  \bhat\Rr_{200}^{-1}$. We display the eigenvalues, as well as the eigenvalue ratios $\lambda_{i}/\lambda_{i+1}$ associated with $\bhat\bXi_{200}$ in Figure \ref{fig:prostate_eigenvalue_gaps}. From the plots of ordered eigenvalues and consecutive eigenvalue ratios, we conclude that the largest 3 eigenvalues of $\bhat\bXi_{200}$ have a clear separation to the rest. Since the third eigenvalue ratio satisfies that $\delta(3) \gg 1.5 \cdot \max_{i\neq 3,\, i\leq 100} \delta(i)$, we conclude $\bhat s = 3$.  

\begin{figure}[h]
	\centering
	\includegraphics[width = \linewidth]{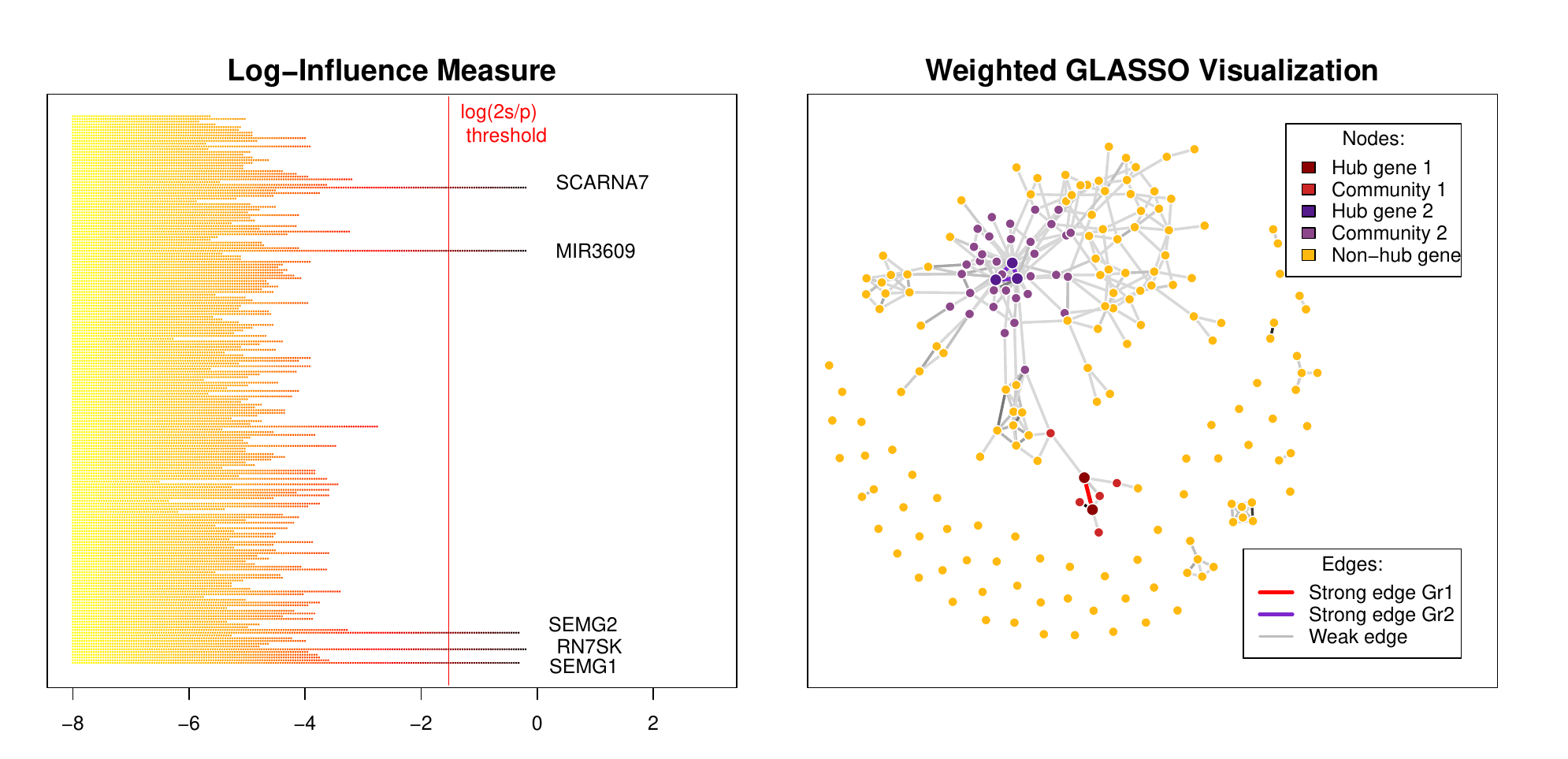}
	\caption{ Outputs of the IPC-HD and 2-step weighted GLASSO methods. Left panel: plot of the log-influence measures on the reduced prostate cancer dataset. A threshold $\kappa = {2\cdot \bhat s}/{p}$ is used. Only five genes are over the threshold: \texttt{SCARNA7}, \texttt{MIR3609}, \texttt{SEMG2}, \texttt{RN7SK}, and \texttt{SEMG1}. Right panel: visualization of the network obtained by the second-step weighted GLASSO. Hub genes are divided in two groups: \{\texttt{SEMG1}, \texttt{SEMG2}\} (red), and \{\texttt{SCARNA7}, \texttt{MIR3609}, \texttt{RN7SK}\} (purple). The former hub group with two hubs is moderately connected to five other genes. The latter group is the center of a community with 37 genes.}
	\label{fig:prosate_network_visualization}
\end{figure}

With $\bhat s = 3$, we proceed to find the influence measures, which are visualized on the left panel of Figure \ref{fig:prosate_network_visualization}. To classify between hubs and non-hubs, we select the threshold $\kappa = {2\cdot \bhat s}/{p} = 0.03$.  Only 5 genes have a measure above the given threshold. These are \texttt{SCARNA7}, \texttt{MIR3609}, \texttt{SEMG2}, \texttt{RN7SK}, and \texttt{SEMG1}. Furthermore, the separation between the influence measures of the estimated hubs and non-hubs is strong. Therefore, there are 5 estimated hub genes in this graphical model. 

To further verify the hub genes detected, we apply to our dataset the 2-step weighted GLASSO method \citep{McGillivray_2020}, which performs graphical model estimation given a previously estimated hub set. Based on the 2-step weighted GLASSO estimator $\bhat\bXi_{\STEP}\in \pdset{200}$ obtained, there are 3 different magnitudes found on the non-diagonal entries. Approximately 73.1\% of all non-diagonal entries in the matrix are zero. Among the non-zero entries, four edges have an absolute value in the rage between $350$ and $870$ units. We  consider these to be strong edges in the graphical model. The strong edges are: $(\texttt{SEMG1},\texttt{SEMG2})$, $(\texttt{RN7SK},\texttt{MIR3609})$, $(\texttt{RN7SK},\texttt{SCARNA7})$, and $(\texttt{MIR3609},\texttt{SCARNA7})$. The rest of the non-zero entries in the matrix range between $10\times 10^{-5}$ and $30$ units. We display the corresponding graphical model, excluding edges with absolute magnitude lower than $1$, on the right panel of Figure \ref{fig:prosate_network_visualization}. From Figure \ref{fig:prosate_network_visualization}, we observe that the five hubs separate into two distinctive groups. The first group of hub genes consists of the pair $\{\texttt{SEMG1},\texttt{SEMG2}\}$ which are connected by a strong edge, and have moderate connections with five non-hub genes. The second group of hub genes includes $\{\texttt{RN7SK},\texttt{SCARNA7},\texttt{MIR3609}\}$, among which all pairwise connections are strong edges. These hub genes are at the center of a cluster of genes, containing 37 genes in total.

Next we discuss the biological role of the hub genes identified by our method, and their relationship to prostate cancer. These genes include \texttt{SEMG1}, \texttt{SEMG2},  \texttt{RN7SK}, and \texttt{MIR3609}. The genes semenogelin I and II (\texttt{SEMG1} and \texttt{SEMG2}) are known for their involvement in the production of seminal-related proteins and it has been shown that their expression levels serve as a prognosis variable of cancer progression \citep{canacci2011expression}. The gene \texttt{RN7SK} is related to cell aging, and it has been found to be under-expressed in stem cells, and human tumor tissues \citep{abasi20167sk}. Finally, the gene \texttt{MIR3609} is related to the level of chemo-resistance of breast cancer cells, and the modification of the expression levels for \texttt{MIR3609} has been explored as a potential method to improve the effectiveness of treatments for prostate cancer \citep{fan2016inhibition}. 

Our identification of these highly relevant hub genes in the prostate cancer dataset can be of interest to practitioners for understanding the underlying process for the development of prostate cancer, as well as potential pathways useful for new treatments.

\section{Discussion}\label{section:discussion}

In this paper, we propose the IPC-HD method for hub detection in GGMs. We achieve this by establishing a connection between the presence of weighted hubs in GGMs, and the leading eigenvectors of the precision matrix. This allow us to detect a set of hubs without requiring the estimation of the entire graphical model. Our IPC-HD method recovers hubs in high-dimensional GGMs when applied to a consistent covariance matrix estimator. 

We provide theoretical guarantees for the IPC-HD, when applied to screened, masked and thresholded covariance matrix estimators, which confirm the effectiveness and reliability of our method even for high-dimensional datasets. Comparative simulations show that our screened and thresholded IPC-HD outperform several existing methods in the literature focusing on the full graph estimation, such as the GLASSO \citep{friedman2008sparse}, the HGL \citep{tan2014learning} and the HWGL \citep{McGillivray_2020}. These simulation results confirm the advantage of the IPC-HD as a direct hub estimation approach. Furthermore, implementing the IPC-HD only requires estimating a covariance matrix, and its spectral decomposition from which the IPC-HD can be implemented efficiently. 

In the future, we aim to study the potential connections between the presence of hubs in graphical models and spectral properties beyond the normality assumption. While studies exist on copula graphical models \citep{liu2012high}, count data \citep{yang2013poisson}, and binary data \citep{ravikumar2010high,tandon2014learning}, it is worth investigating whether the proposed spectral method can be extended to these diverse contexts.

\subsection*{Supplementary Materials}

We provide proofs, additional discussions and simulations in the Supplementary Materials. Code and data can be found in our Github repository: \begin{center}\url{https://github.com/jsgomez94/HubEstimationCodeSubmission.git}.\end{center}

\setstretch{1}
\bibliographystyle{apalike}
\bibliography{ref}

\end{document}